\documentclass[aps, prb, twocolumn, superscriptaddress, longbibliography, showkeys]{revtex4-1}
\usepackage[colorlinks, linkcolor=blue, anchorcolor=blue, citecolor=blue]{hyperref}
\usepackage{amsmath}
\usepackage{graphicx}
\usepackage{braket}
\usepackage{multirow}
\usepackage{color}
\usepackage{soul}
\usepackage{diagbox}
\usepackage{rotating}

\usepackage{ulem}

\begin{document}

\title{Electronic Correlation Effects on Stabilizing a Perfect Kagome Lattice and Ferromagnetic Fluctuation in LaRu$_3$Si$_2$}
\author{Yilin Wang}    
\affiliation{Hefei National Laboratory for Physical Sciences at Microscale, University of Science and Technology of China, Hefei, Anhui 230026, China} 

\date{\today}

\begin{abstract}  
A perfect Kagome lattice features flat bands that usually lead to strong electronic correlation effects, but how electronic correlation, in turn, stabilizes a perfect Kagome lattice has rarely been explored. Here, we study such effect in a superconducting ($T_c \sim 7.8$ K) Kagome metal LaRu$_3$Si$_2$ with a distorted Kagome plane consisting of pure Ru ions, using density functional theory plus $U$ and plus dynamical mean-field theory. We find that increasing electronic correlation can stabilize a perfect Kagome lattice and induce substantial ferromagnetic fluctuations in LaRu$_3$Si$_2$. By comparing the calculated magnetic susceptibilities to experimental data, LaRu$_3$Si$_2$ is found to be on the verge of becoming a perfect Kagome lattice. It thus shows moderate but non-negligible electronic correlations and ferromagnetic fluctuations, which are crucial to understanding the experimentally observed non-Fermi-liquid behavior and the pretty high superconducting $T_c$ of LaRu$_3$Si$_2$. 
\end{abstract}

\keywords{Electronic Correlation, Kagome lattice, LaRu$_3$Si$_2$, Ferromagnetic Fluctuation, Flat bands}

\maketitle

\section{Introduction} 

A perfect Kagome lattice features flat bands due to the destructive interference of electron hoppings within the geometrically frustrated structure~\cite{syozi:1951,Mielke:1991,Tasaki:1992,Sachdev:1992,wenxg:2011}. When flat bands are near the Fermi level, it usually induces strong electronic correlation effects such as unconventional superconductivity~\cite{Cao:2018,Balents:2020,Aoki:2020,Heikkila:2016,JiangKun:2022,Nie:2022}, magnetism~\cite{Mielke:1991,Tasaki:1992}, topological phases~\cite{wenxg:2011} and exotic charge density waves~\cite{Jiang:2021,Teng:2022}. In recent years, these flat-bands-induced correlation effects have been widely studied in Kagome metals, for example, FeSn~\cite{Kang:2020,Zhenyu:2020}, FeGe~\cite{Huang:2020,Teng:2022}, Fe$_3$Sn$_2$~\cite{Ye:2018,Zhenyu:2018,Yin:2018}, CoSn~\cite{Kang:2020ER,Liu:2020,Yin:2020,Brian:2020,Changgan:2022}, RT$_6$Ge$_6$ (R=rare-earth elements, T=Mn, Cr)~\cite{Yin:2020b,Yang:2022}. But how electronic correlation, in turn, stabilizes a perfect Kagome lattice has rarely been explored. Here, we study such effect in a Kagome metal LaRu$_3$Si$_2$~\cite{Barz:1980,Haihu:2016,Haihu:2011,Mielke:2021,Gong:2022}, using density functional theory (DFT) plus $U$~\cite{Anisimov:1991} and plus dynamical mean-field theory (DMFT)~\cite{Georges:1996,lichtenstein:2001,kotliar:2006}.

LaRu$_3$Si$_2$ is a superconductor with a highest $T_c\sim 7.8$ K~\cite{Barz:1980} among the known Kagome superconductors at ambient conditions. It is a paramagnetic metal at high-temperate. Non-Fermi-liquid (NFL) behavior was inferred from transport experiments, indicating substantial electronic correlations from Ru-$4d$ orbitals~\cite{Haihu:2011}. A recent DFT calculation based on the electron-phonon coupling mechanism yields a $T_c\sim 1.2$ K, much smaller than the experimental value~\cite{Mielke:2021}, also indicating other factors such as electronic correlations and magnetic fluctuations must be considered to understand its superconductivity.

\begin{figure}
    \centering
    \includegraphics[width=0.5\textwidth]{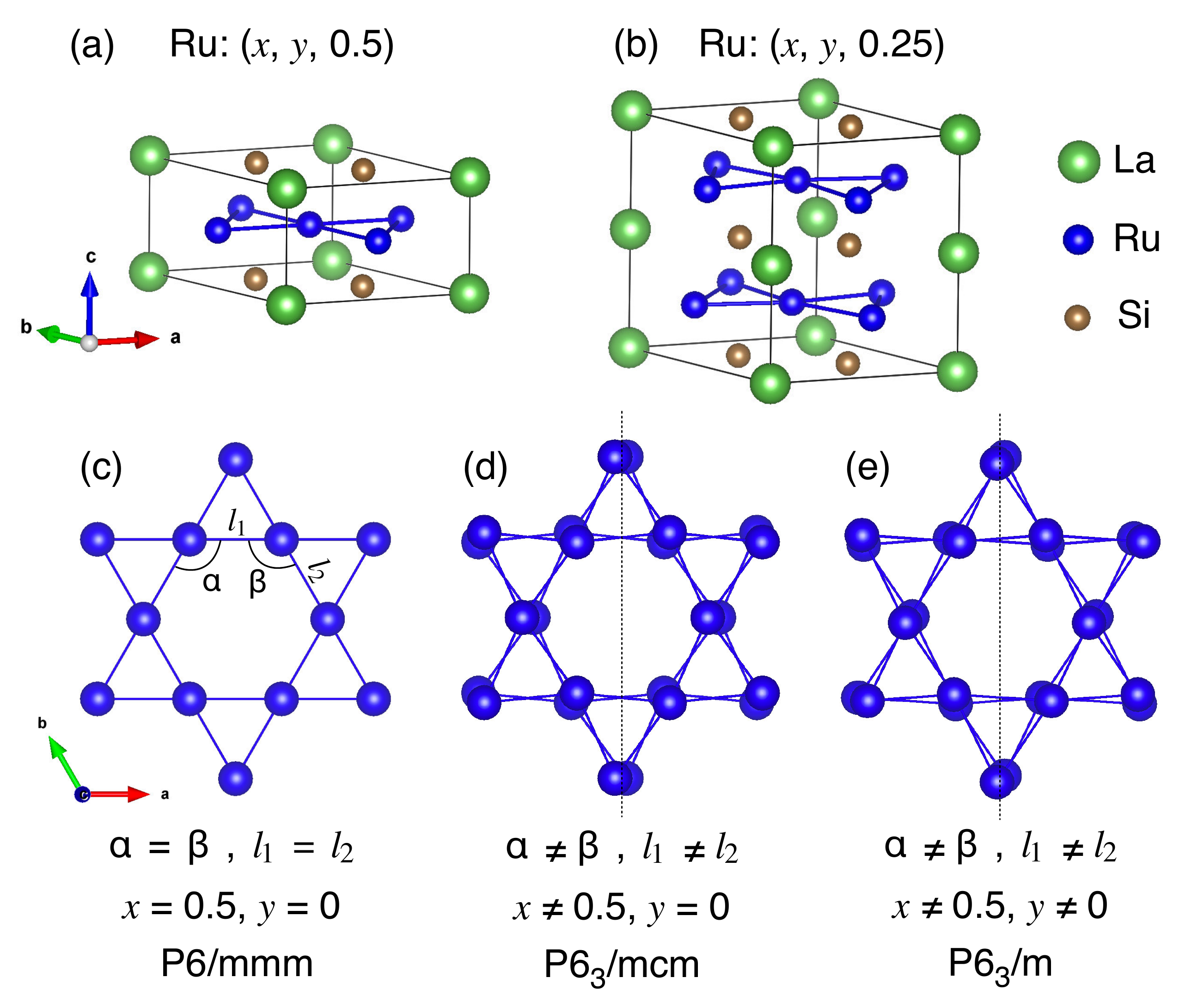}
    \caption{Three possible crystal structures of LaRu$_3$Si$_2$ with a Kagome plane consisting of pure Ru atoms. (a),(c) Perfect Kagome structure. (b),(d),(e) Two possible distorted Kagome structures with a doubling of the $c$-axis. (c)-(e) Top view of the Kagome planes. The angles $\alpha,\beta$ and side lengths $l_1,l_2$ of the hexagon, the fractional coordinates of Ru atoms $x,y$, and the space group are shown, respectively. The crystal structures are constructed by VESTA~\cite{VESTA:2008}. }
    \label{fig:struct}
\end{figure}

The perfect Kagome structure of LaRu$_3$Si$_2$ with space group P6/mmm is shown in Fig.~\ref{fig:struct}(a) and \ref{fig:struct}(c). The Kagome layer consists of pure Ru atoms, different from most other Kagome materials in which an anion usually resides in the center of the hexagon of the Kagome structure. Adjacent to the Kagome layer is a layer consisting of a triangular lattice of La and a honeycomb lattice of Si. However, a previous X-Ray diffraction study shows that LaRu$_3$Si$_2$, in reality, crystallizes into a slightly distorted Kagome structure with a doubling of the $c$-axis and the space group P6$_3$/m~\cite{Barz:1980} (see Fig.~\ref{fig:struct}(b) and \ref{fig:struct}(e)). Given only one kind of ion (Ru) present in the Kagome layer of LaRu$_3$Si$_2$, it provides an ideal platform for studying its electronic correlation effects on stabilizing a perfect Ru Kagome lattice, since it rules out possible crystalline field effects of anion on the stability of the Kagome lattice.

Our calculations find that increasing electronic correlation can stabilize a perfect Kagome lattice and induce substantial ferromagnetic fluctuations in LaRu$_3$Si$_2$. By comparing the calculated magnetic susceptibilities to experimental data, LaRu$_3$Si$_2$ is found to be on the verge of the transition from a distorted to a perfect Kagome lattice. It thus shows moderate but non-negligible electronic correlations and ferromagnetic fluctuations, consistent with the experimentally observed NFL behavior~\cite{Haihu:2011}. 
Furthermore, our calculations show that the distorted Kagome structure of LaRu$_3$Si$_2$ may hold higher symmetry (space group P6$_3$/mcm, see Fig.~\ref{fig:struct}(d)) than that reported by previous experiment (P6$_3$/m)~\cite{Barz:1980}, which should be further examined by high-resolution crystal structure refinement with high-quality samples.

\section{Methods} 

We perform DFT+U calculations using the VASP package~\cite{kresse:1996,blochl:1994}, with exchange-correlation functional of both local density approximation (LDA) and generalized gradient approximation (GGA)~\cite{perdew:1996}. The energy cutoff of the plane-wave basis is set to be 500 eV, and a $\Gamma$-centered $21\times 21\times 21$ K-point grid is used. The internal atomic positions are relaxed in the non-magnetic states, until the force of each atom is smaller than 1 meV/\AA. The rotationally invariant DFT+U method introduced by Liechtenstein \textit{et al.}~\cite{Liechtenstein:1995} is used, which is parameterized by Hubbard $U$ and Hund's coupling $J_H$ (LDAUTYPE=1 or 4). It turns out that the spin-orbital coupling (SOC) of Ru would not change the main conclusions, so we only present the non-SOC results in the main text, and a SOC result is presented in the Supporting Information (Fig. S1)~\cite{suppl}. The energies of two magnetic orders, ferromagnetic (FM) and A-type anti-ferromagnetic (AFM) (Fig. S2)~\cite{suppl}, are also calculated by GGA+U.

We also perform fully charge self-consistent LDA+DMFT calculations in the paramagnetic states of LaRu$_3$Si$_2$, using the code EDMFTF developed by Haule \textit{et al.}~\cite{Haule:2010,Haule:2015free} based on the WIEN2K package~\cite{Blaha:2020}. We choose a wide hybridization energy window from -10 to 10 eV with respect to the Fermi level. All five Ru-$4d$ orbitals are considered as correlated ones and a local Coulomb interaction Hamiltonian with rotationally invariant form is applied. The local Anderson impurity model is solved by the continuous time quantum Monte Carlo (CTQMC) solver~\cite{Gull:2011}. We use an “exact” double counting scheme invented by Haule~\cite{Haule:2015}. The self-energy on real frequency $\Sigma(\omega)$ is obtained by the analytical continuation method of maximum entropy. We follow the method introduced by Haule \textit{et al.}~\cite{Haule:2016force} to perform structure relaxation in the framework of LDA+DMFT. All the calculations are preformed at $T=290$ K. Following Ref.~\cite{Haule:2015free}, we use the Yukawa representation of the screened Coulomb interaction, in which there is an unique relationship between $U$ and $J_H$. If $U$ is specified, $J_H$ is uniquely determined by a code in EDMFTF~\cite{tutor:MnO}. Their values are tabulated in the Supporting Information (Table S1)~\cite{suppl} and are also used for the DFT+U calculations.

\begin{figure}
    \centering
    \includegraphics[width=0.5\textwidth]{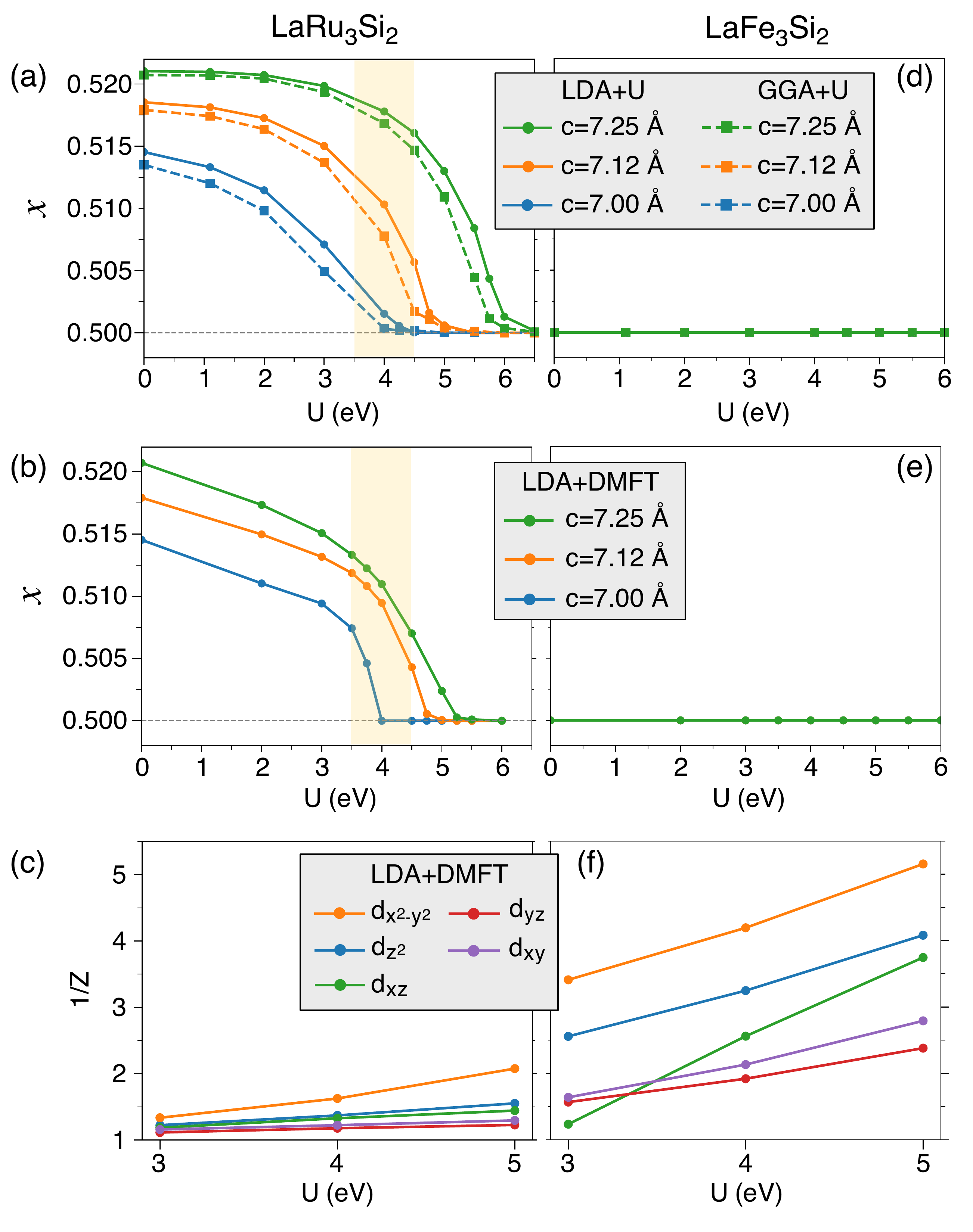}
    \caption{Electronic correlation effects on stabilizing a perfect Kagome lattice. (a),(d) Fractional coordinates $x$ of Ru (Fe) relaxed by LDA+$U$ and GGA+$U$ as functions of Hubbard $U$, for different lattice parameter ratio $c/a$ with the fixed crystal volume of experiment. $c=7.12$ \AA \ is the experimental value. $x=0.5$ for a perfect Kagome lattice. (b),(e) $x$ relaxed by LDA+DMFT as functions of $U$. The yellow area mark that LaRu$_3$Si$_2$ is on the verge of becoming a perfect Kagome lattice. (c),(f) LDA+DMFT calculated mass-enhancement of Ru-$4d$ (Fe-$3d$) orbitals due to electronic correlations, as functions of $U$. (a)-(c) for LaRu$_3$Si$_2$ and (d)-(f) for LaFe$_3$Si$_2$.}
    \label{fig:relax}
\end{figure}

To reveal the role of electronic correlation effects on stabilizing a perfect Kagome lattice in LaRu$_3$Si$_2$, (1) we vary the Hubbard $U$ from 0 to 6 eV in our calculation; 
(2) we perform a comparative study on a hypothetical crystal structure LaFe$_3$Si$_2$ with the same lattice parameters and initial atomic positions as LaRu$_3$Si$_2$, since Fe-$3d$ orbitals are expected to show stronger electronic correlations than Ru-$4d$ orbitals. To uncover how the adjacent LaSi$_2$ layers affect the stability of the Ru$_3$ Kagome layer, 
we also perform comparative calculations by varying the lattice parameter ratio $c/a$ with the fixed crystal volume from experiment. The experimental lattice parameters are $a=5.676 $ \AA, $c=7.12$ \AA, which gives a volume of 198.65 \AA$^3$~\cite{Barz:1980}. Crystal structure with space group P6$_3$/m and Ru sites at ($x=0.52$, $y=0.01$, $z=0.25$) is constructed as the starting point for relaxation (see Fig.~\ref{fig:struct}(b) and \ref{fig:struct}(e)). However, all the relaxations converge to the higher symmetry structure with space group P6$_3$/mcm ($x\ne 0.5$, $y=0$, $z=0.25$), where an additional $C_2$ and mirror symmetry is present (see Fig.~\ref{fig:struct}(d)). Therefore, we will only discuss the results of space group P6$_3$/mcm below.
    
\section{Results} 

Fig.~\ref{fig:relax}(a)-(b) and (d)-(e) show the fractional coordinate $x$ of Ru (Fe) relaxed by LDA+U, GGA+U and LDA+DMFT methods, for different $c/a$ ratio. For LaRu$_3$Si$_2$, as increasing $U$, all the methods yield a tendency that $x$ is decreasing and approaching to the value of a perfect Kagome lattice ($=0.5$). We note that LaFe$_3$Si$_2$ converges to the perfect Kagome structure even at $U=0$. Comparing to LDA+U, the GGA+U method that is expected to better describe the correlation effects, gives smaller $x$ (dashed curves in Fig.~\ref{fig:relax}(a)). Fig.~\ref{fig:relax}(c) and (f) show the LDA+DMFT calculated orbital-resolved quasi-particle mass-enhancement, $m^\star/m^{DFT}=1/Z$, due to the electronic correlation effects, where $Z$ is quasi-particle weight. The system is more correlated as $1/Z$ deviates more from 1. The electronic correlation become stronger as increasing $U$, and LaFe$_3$Si$_2$ shows much stronger correlation than LaRu$_3$Si$_2$ as expected. The most correlated orbital is $d_{x^2-y^2}$ which contributes to the flat bands near the Fermi level (see below). Therefore, these results indicate that strong electronic correlation would stabilize a perfect Kagome lattice consisting of transition metal ions in the system of LaX$_3$Si$_2$ (X=Ru, Fe).

Fig.~\ref{fig:relax}(a)-(b) also show that increasing the lattice parameter ratio $c/a$, i.e., longer distance between the Ru$_3$ Kagome layer and LaSi$_2$ layer, tends to increase $x$ and distort the Kagome plane. Thus, larger Hubbard $U$ is required to stabilize a perfect Kagome lattice at larger ratio of $c/a$.

\begin{figure}
    \centering
    \includegraphics[width=0.5\textwidth]{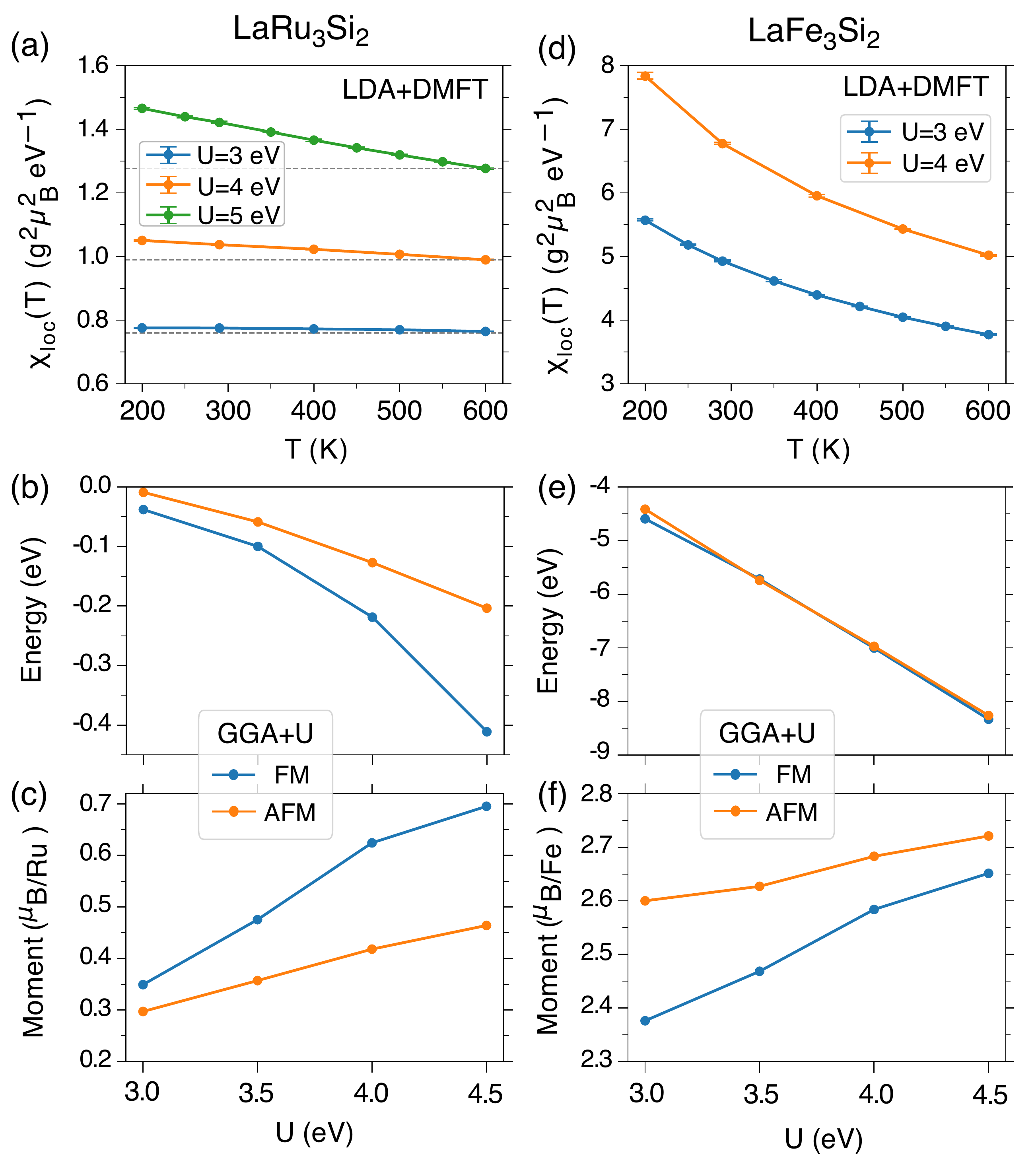}
    \caption{Electronic correlation induced magnetism. (a),(d) Local magnetic susceptibilities calculated by LDA+DMFT as functions of temperate $T$ for different $U$ values. (b),(e) The energies (per unit cell) of FM and AFM orders with respect to the non-magnetic state as functions of $U$, calculated by GGA+$U$. (c),(f) Ordered magnetic moment in the FM and AFM states. (a)-(c) for LaRu$_3$Si$_2$, the crystal structures relaxed at the corresponding $U$ are used. (d)-(f) for LaFe$_3$Si$_2$. The experimental lattice parameters, $a=5.676$\AA\; and $c=7.12$\AA, are used for both compounds.}
    \label{fig:mag}
\end{figure}

Fig.~\ref{fig:mag}(a) and (d) show the local magnetic susceptibilities $\chi_{\text{loc}}(T)$ calculated by LDA+DMFT for different $U$. For LaRu$_3$Si$_2$ , $\chi_{\text{loc}}(T)$ shows Pauli paramagnetism behavior at $U=3$ eV, while it shows a Curie-Weiss-like behavior at $U=5$ eV. In between, $\chi_{\text{loc}}(T)$ slightly increases as decreasing temperate, which is consistent with that measured by experiment (see Fig.7 in Ref.~\cite{Haihu:2011}). Based on this, we infer that the actual Hubbard $U$ for LaRu$_3$Si$_2$ is about 4 eV. At this $U$ value, LaRu$_3$Si$_2$ is thus found to be on the verge of crystallizing into a perfect Kagome structure, according to the relaxation results at the experimental lattice parameters shown as the orange curves in Fig.~\ref{fig:relax}(a),(b). At $U=4$ eV, the mass-enhancement of the most correlated orbital $d_{x^2-y^2}$ is about 1.63 (Fig.~\ref{fig:relax}(c)). This indicates a moderate electronic correlation in LaRu$_3$Si$_2$, consistent with a slightly large Wilson ratio $R=2.88$ ($R=1$ for non-interacting electron gas) and a NFL contribution to the electronic specific heat, found by the transport experiment~\cite{Haihu:2011}. In contrast, $\chi_{\text{loc}}(T)$ shows a well-defined Curie-Weiss behavior for LaFe$_3$Si$_2$, indicating much stronger electronic correlation and magnetism.

\begin{figure*}
    \centering
    \includegraphics[width=0.9\textwidth]{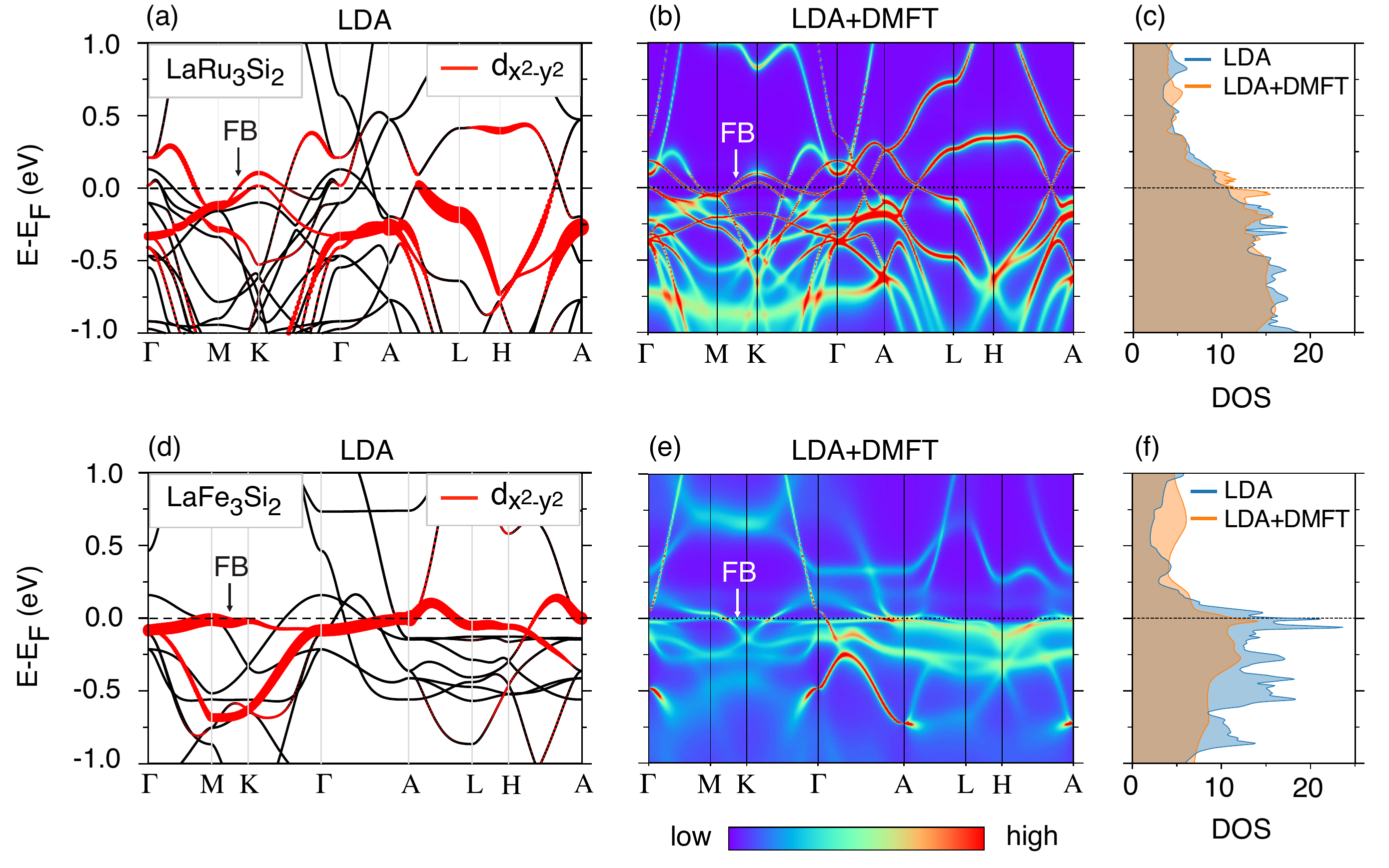}
    \caption{Flat band near Fermi level. (a),(d) LDA calculated band structures. The flat bands (FB) with $d_{x^2-y^2}$ character are shown in red. (b),(e) LDA+DMFT calculated spectrum function $A(k,\omega)$ at $U=4$ eV, $J_H=0.782$ eV and $T=290$ K. (c),(f) The corresponding density of states. (a)-(c) for LaRu$_3$Si$_2$, the crystal structure relaxed by LDA+DMFT at $U=4$ eV, $J_H=0.782$ eV is used. (d)-(f) for LaFe$_3$Si$_2$. The experimental lattice parameters, $a=5.676$\AA\; and $c=7.12$\AA, are used for both compounds.}
    \label{fig:band}
\end{figure*}

Fig.~\ref{fig:mag}(b) and (e) show the energies of the FM and AFM orders with respect to the non-magnetic phase calculated by GGA+U, and Fig.~\ref{fig:mag}(c) and (f) show the corresponding ordered magnetic moments. LaFe$_3$Si$_2$ has very low energies in magnetic states compared to its non-magnetic state and large ordered magnetic moments, suggesting that it tends to order as decreasing temperature. Its FM and AFM states are very close in energies, indicating comparable inter-layer FM and AFM couplings. LaRu$_3$Si$_2$ shows much weaker magnetism compared to LaFe$_3$Si$_2$, but the FM and AFM states are still found to be energetically stable at the static mean-field level. Its FM state has significantly lower energy than the AFM state. Although no magnetic orders have been observed experimentally in LaRu$_3$Si$_2$, the stable FM state found by GGA+U calculation suggests that substantial FM fluctuations may exist in LaRu$_3$Si$_2$. Such FM fluctuations should be attributed to the flat bands near the Fermi level, derived from the Kagome lattice.

Fig.~\ref{fig:band} shows the LDA and LDA+DMFT calculated band structures and density of states (DOS) at $U=4$ eV and $J_H=0.782$ eV in the paramagnetic states.
Extremely flat bands with $d_{x^2-y^2}$ characters clearly exist near the Fermi level in LaFe$_3$Si$_2$. However, such bands in LaRu$_3$Si$_2$ are not as flat as that in LaFe$_3$Si$_2$. By comparing the band structures between the distorted and perfect Kagome lattice of LaRu$_3$Si$_2$, we found that the magnitude of distortion of the Kagome lattice have small effects on the flat bands in LaRu$_3$Si$_2$. Actually, it is mainly caused by the more extended $4d$ orbitals that will induce non-vanishing hoppings among distant Kagome lattice sites and result in imperfect destructive interference of hoppings.
In LaRu$_3$Si$_2$, the flat bands would induce substantial FM fluctuations, as shown in Fig.~\ref{fig:mag}(b).

\section{Conclusion and discussion} 

To summarize, by taking the Kagome system LaX$_3$Si$_2$ (X=Ru, Fe) as an example, we have demonstrated that strong electronic correlations could play important roles in stabilizing a perfect Kagome plane. We show that LaRu$_3$Si$_2$ is on the verge of becoming a perfect Kagome lattice and it thus exhibits moderate electronic correlation and substantial ferromagnetic fluctuations. 
Previous study has shown that the electron-phonon couplings alone are not nearly enough to account for a superconducting $T_c$ of 7.8 K in LaRu$_3$Si$_2$~\cite{Mielke:2021}. The electronic correlations and ferromagnetic fluctuations found in our study may play significant roles in enhancing $T_c$ of LaRu$_3$Si$_2$, which is worthy of further study.

As shown in the Fig. S3 in the Supporting Information~\cite{suppl}, the only obvious difference in the simulated powder diffraction pattern of LaRu$_3$Si$_2$ between the space group P6$_3$/mcm and P6$_3$/m is an additional reflection at (1 0 1) for P6$_3$/m. However, its intensity is very weak compared to the main peak such that it is very difficult to be resolved experimentally. Further high-resolution crystal structure refinement with high-quality sample is required to determine whether LaRu$_3$Si$_2$ crystallizes into the space group P6$_3$/mcm or not. Our result thus provides a reference for the crystal structure refinement of LaRu$_3$Si$_2$.



\section*{Acknowledgment}
This work was supported by USTC Research Funds of the Double First-Class Initiative (No. YD2340002005). All the calculations presented in this work were preformed on TianHe-1A, the National Supercomputer Center in Tianjin, China.  



\bibliography{main}

\end{document}


\title{Supporting Information for ``Electronic Correlation Effects on Stabilizing a Perfect Kagome Lattice and Ferromagnetic Fluctuation in LaRu$_3$Si$_2$''}
\author{Yilin Wang}     
\affiliation{Hefei National Laboratory for Physical Sciences at Microscale, University of Science and Technology of China, Hefei, Anhui 230026, China} 

\date{\today}

\begin{abstract}
\end{abstract}

\maketitle


\begin{table*}
    \centering
    \caption{Values of Hubbard $U$ and Hund's coupling $J_H$ calculated by the code \emph{R\underline{ }Coulomb.py} in DFT+EDMFTF package. These values are used for both DFT+U and LDA+DMFT calculations.}
    \begin{ruledtabular}
    \begin{tabular}{cccccccccc}
    $U$ (eV)   & 1.1   & 1.5   & 2.0   & 3.0   & 4.0   & 4.5  & 5.0   & 5.5   & 6.0\\
    $J_H$ (eV) & 0.389 & 0.476 & 0.563 & 0.692 & 0.782 &0.817 & 0.848 & 0.874 & 0.897\\
    \end{tabular}
    \end{ruledtabular}
    \label{tab:multi}
    \end{table*}

\begin{figure*}
        \centering
        \includegraphics[width=0.9\textwidth]{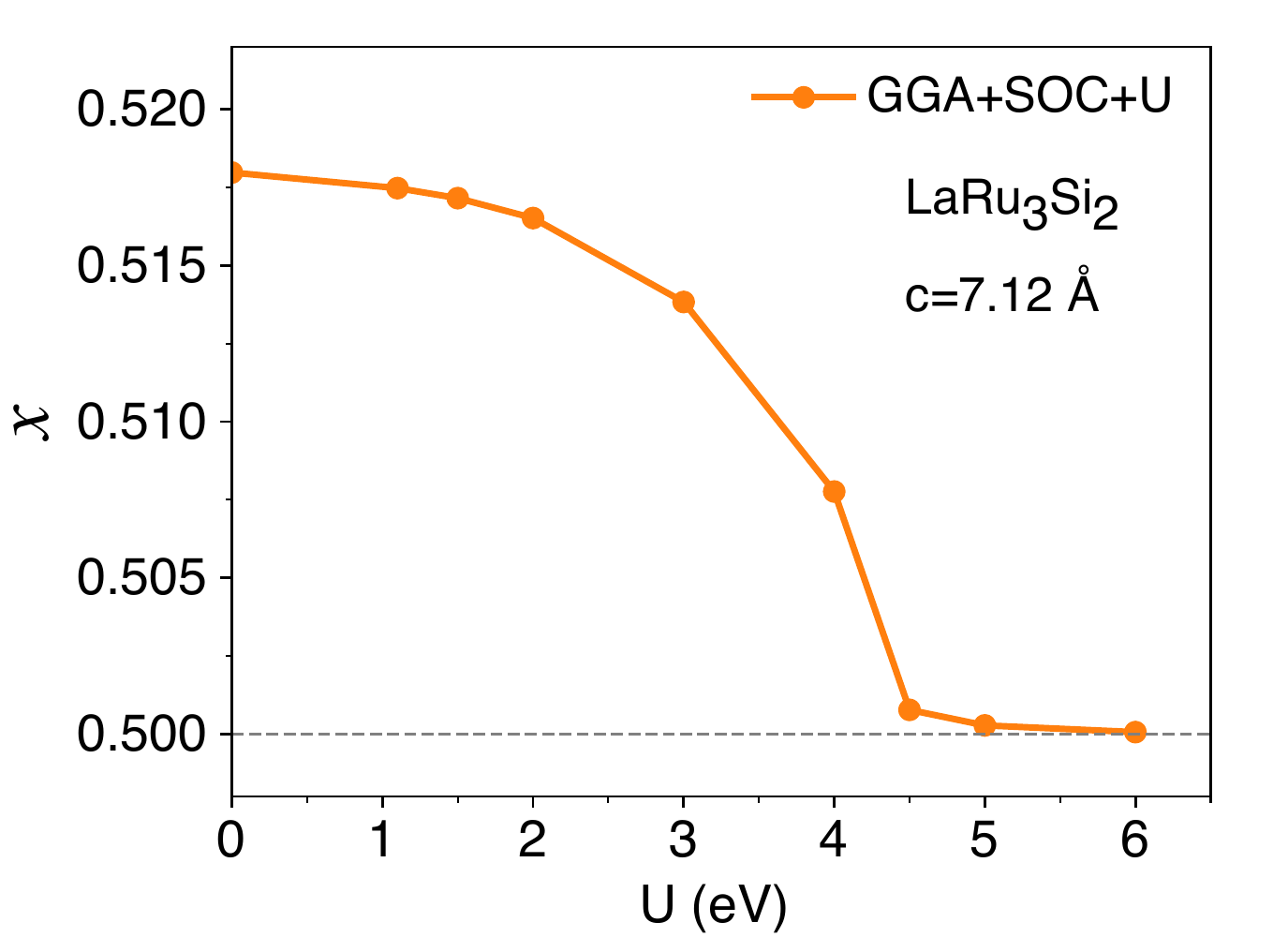}
        \caption{Fractional coordinates $x$ of Ru sites as function of Hubbard $U$, relaxed by GGA+U with spin-orbital coupling.}
        \label{fig:ggasoc}
\end{figure*}

\begin{figure*}
    \centering
    \includegraphics[width=0.9\textwidth]{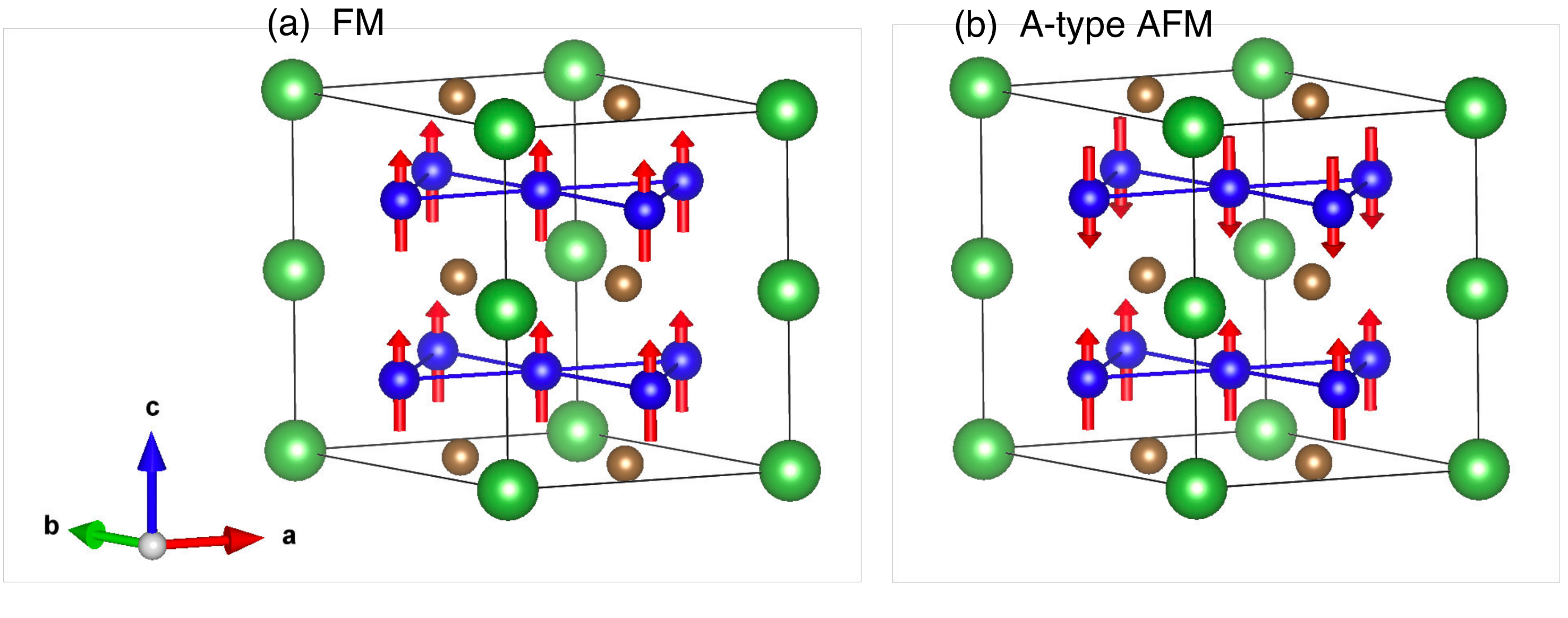}
    \caption{Magnetic configurations considered in the GGA+U calculations.}
    \label{fig:mag_conf}
\end{figure*}

\begin{figure*}
    \centering
    \includegraphics[width=0.9\textwidth]{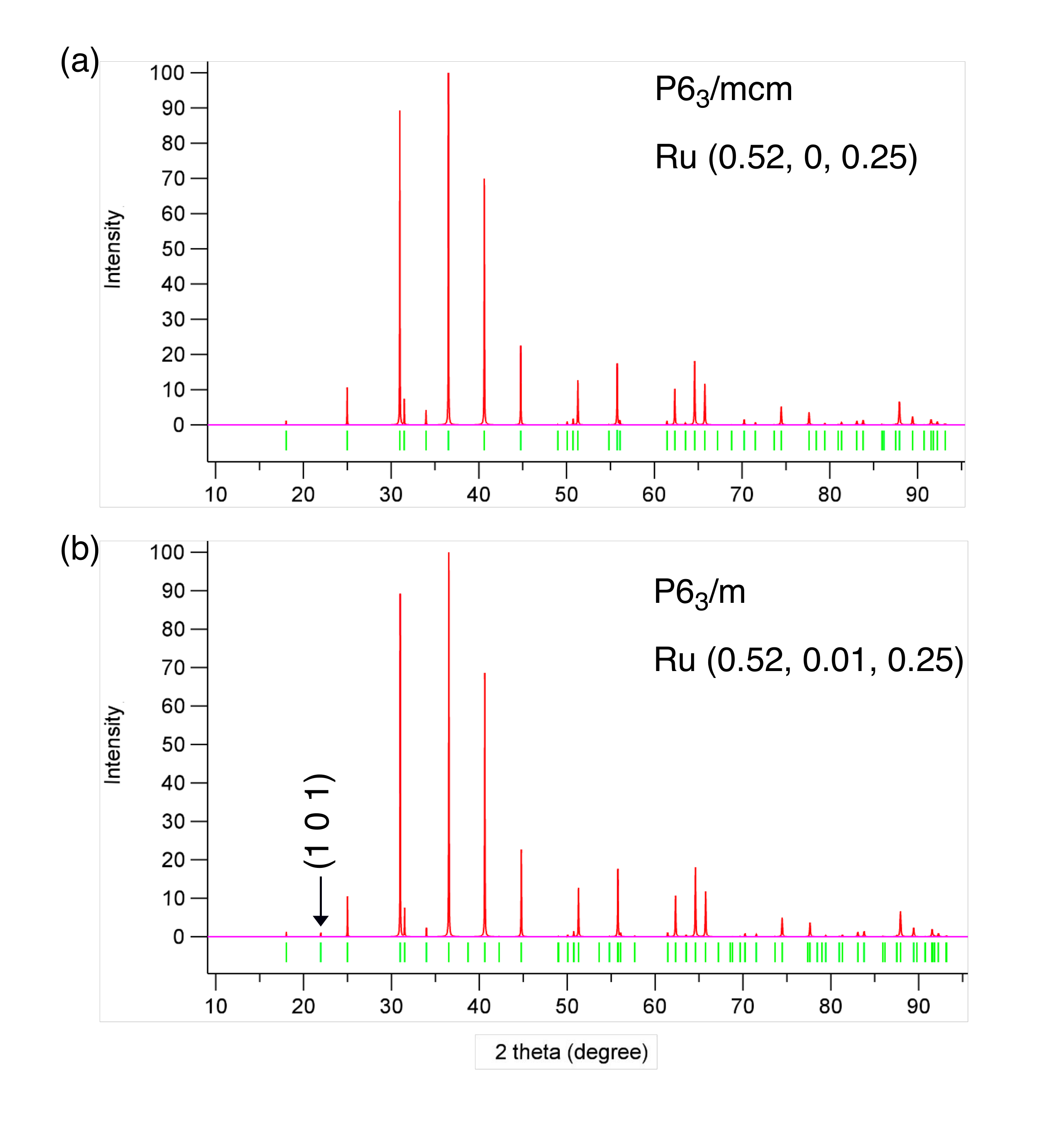}
    \caption{Simulated XRD pattern of the possible distorted Kagome structure of LaRu$_3$Si$_2$. (a) For space group P6$_3$/mcm with Ru at (0.52, 0, 0.25). (b) For space group P6$_3$/m with Ru at (0.52, 0.01, 0.25). Lattice parameters are $a=5.676$\AA\ and $c=7.12$\AA. Their only difference is that there is an additional weak peak at (1 0 1) for P6$_3$/m.  }
    \label{fig:xrd}
\end{figure*}

\bibliography{suppl}